\documentclass{aa}  

\usepackage{graphicx}
\usepackage{txfonts}
\newcommand\kms{\ifmmode{\rm km\thinspace s^{-1}}\else km\thinspace s$^{-1}$\fi}
\newcommand\nsvs{NSVS~10653195}

\begin{document}

   \title{Absolute dimensions of the low-mass eclipsing binary system \nsvs.
   \thanks{Tables 1, 2, 3 and 7 are only available in electronic form
at the CDS via anonymous ftp to cdsarc.u-strasbg.fr (130.79.128.5)
or via http://cdsweb.u-strasbg.fr/cgi-bin/qcat?J/A+A/}}
   

   \author{Ram\'on Iglesias-Marzoa\inst{1,2}
          \and
          Mar\'ia J. Ar\'evalo\inst{1,3}
          \and          
          Mercedes L\'opez-Morales\inst{4}
          \and
          Guillermo Torres\inst{4}
          \and
          Carlos L\'azaro\inst{1,3}
          \and
          Jeffrey L. Coughlin\inst{5}
          }

   \institute{Astrophysics Department, Universidad de La Laguna,
              E-38205 La Laguna, Tenerife, Spain
              \and
             Centro de Estudios de F\'isica del Cosmos de Arag\'on,
             Plaza San Juan 1, E-44001, Teruel, Spain\\
             \email{riglesias@cefca.es} 
             \and
             Instituto de Astrof\'isica de Canarias,
             E-38200, La Laguna, Tenerife, Spain
             \and
             Center for Astrophysics \textbar\ Harvard \& Smithsonian,
             60 Garden Street, Cambridge, MA 02138, USA
             \and
             NASA AMES Research Center, Moffet Field, CA, USA
             }

   \date{Received March 22, 2019; accepted June 06, 2019}

  \abstract
   {Low-mass stars in eclipsing binary systems show radii larger and effective temperatures lower than theoretical stellar models predict for
   isolated stars with the same masses. Eclipsing binaries with low-mass components are hard to find due to their low luminosity. As a
   consequence, the analysis of the known low-mass eclipsing systems is key to understand this behavior.
   }
   {We aim to investigate the mass--radius relation for low-mass stars and the cause of the deviation
   of the observed radii in low-mass detached eclipsing binary stars (LMDEB) from theoretical stellar models.}
   {We developed a physical model of the LMDEB system \nsvs\ to accurately measure
   the masses and radii of the components. We obtained several high-resolution spectra in order to fit a spectroscopic orbit.
   Standardized absolute photometry was obtained to measure reliable color indices and to measure the mean $T_{\rm eff}$
   of the system in out-of-eclipse phases.
   We observed and analyzed optical $VRI$ and infrared $JK$ band differential light-curves which
   were fitted using PHOEBE. A Markov-Chain Monte Carlo (MCMC) simulation near the solution
   found provides robust uncertainties for the fitted parameters.
   }
   {\nsvs\ is a detached eclipsing binary composed of two similar stars with masses of $M_1=0.6402\pm0.0052$ $M_\sun$ and
   $M_2=0.6511\pm0.0052$ $M_\sun$ and radii of $R_1=0.687^{+0.017}_{-0.024}$ $R_\sun$ and $R_2=0.672^{+0.018}_{-0.022}$ $R_\sun$.
   Spectral types were estimated to be K6V and K7V. These stars rotate in a circular orbit with an orbital inclination of $i=86.22\pm0.61$
   degrees and a period of $P=0.5607222(2)$ d. The distance to the system is estimated to be $d=135.2^{+7.6}_{-7.9}$ pc,
   in excellent agreement with the value from Gaia. If solar metallicity were assumed, the age of the system
   would be older than $\log(age)\sim$8 based on the $M_{bol}$-$\log T_{\rm eff}$ diagram.
   }
   {\nsvs\ is composed of two oversized and active K stars. While their radii is above model predictions
   their $T_{\rm eff}$ are in better agreement with models.
   }


   \maketitle
%

\section{Introduction}
\label{sec:Introduction}

Eclipsing binary stars are currently the most powerful tool to simultaneously measure the radii and masses of stars.
By observing a photometric light curve (LC) and the radial velocity (RV) curves of the two components of the binary system
it is possible to measure their masses and radii with precisions of a few percent, depending on the quality
of the observational data. In recent years these techniques have also lead to the characterization of more than 3900
exoplanets around stars other than the Sun (see online databases\footnote{\url{http://exoplanet.eu/catalog/}}
\footnote{\url{https://exoplanetarchive.ipac.caltech.edu/docs/counts_detail.html}}
for updated lists).

Stars in the upper main sequence are intrinsically luminous and can be observed at great distance
in our Galaxy and even in other nearby galaxies \citep[see, e.g.][]{ribas2005, pietrzynski2013, lee2014}.
As a result, applying these techniques to the upper main sequence stars provided a great number of stellar
parameters and distances in that range of masses \citep[see][for a list of well measured systems]{torres2010b}.
But the lower main sequence is a different issue, in particular the so-called low-mass stars,
those with masses below 1 $M_{\sun}$ and spectral types K and M. These stars are intrinsically faint and
until a decade ago only half a dozen low-mass detached eclipsing binary systems (hereafter LMDEB)
in the solar neighborhood were known, all of them with short periods ($\lesssim 1$ d or less).
To make things worse, the low-mass stars are also affected by chromospheric and coronal activity caused by strong
magnetic fields, dark and bright spots, flares, and plages. This activity can be observed as flares and
modulation of the LCs at the out-of-eclipse orbital phases, X-ray emission, and emission spectral lines --
usually CaII H and K or hydrogen lines (e.g., $H_\alpha$). These features are difficult to model and
add additional uncertainties to the physical parameters.

The data provided by the last generation of photometric surveys have improved this situation.
Millions of LCs have thus far been provided by variable star and transient photometric databases like the
North Sky Variability Survey \citep[NSVS, ][]{wozniak2004}, or
the All-Sky Automated Survey \citep[ASAS, ][]{pojmanski1997}, to name only two, databases devoted to searching for exoplanet transits from Earth, for example
Wide Angle Search for Planets \citep[WASP, ][]{pollacco2006},
Hungarian Automated Telescope \citep[HAT, ][]{bakos2004} and others,
as well as databases devoted to searching for exoplanet transits from space, such as
COnvection, ROtation and planetary Transits, \citep[COROT, ][]{fridlund2006ESASP}, or
\textit{Kepler} \citep{borucki2010}. These databases are available to be screened for for
LMDEBs
\citep[see, e.g.,][]{shaw2007, coughlin2011, irwin2011ApJ, hartman2018AJ}
and other variable stars.

The analysis of those first LMDEB systems resulted in two striking features: in the mass--radius (M-R) relation
the radii measured for most stars are greater than the model predictions by 10-15\%, well above the
uncertainties of the parameters. Also, the effective temperatures ($T_{\rm eff}$) of the stars, plotted in a
$T_{\rm eff}$-M diagram, are 5-7\% lower than the models predict \citep{lopez-morales2005, morales2010}.
At present, the main suspect is the activity generated by the strong magnetic fields,
enhanced by the short rotational periods of these stars,
although other alternative explanations were proposed elsewhere \citep{lopez-morales2007}.
Current models do not account for this stellar activity, which is common in these types of stars.
Therefore, it is desirable to analyze more LMDEB systems in order to see whether or not this is a common feature to all
these systems, and to check possible dependences on parameters of the sample such as the orbital period.

\object{\nsvs}\ = 2MASS J16072787+1213590 = SDSS J160727.85+121358.9 ($\alpha$=16:07:27.86, $\delta$=+12:13:59.1, J2000.0)
is an eclipsing binary star first published as a LMDEB candidate in a list by \cite{shaw2007} after a search
for LMDEBs among the variable stars of the North Sky Variability Survey \citep[NSVS,][]{wozniak2004}. 
The first light curve analysis of this system was done by \cite{coughlin2007} who measured
high-precision Johnson $VRI$ light curves and provided the first light-curve analysis using
Eclipsing Light Curve \cite[ELC,][]{orosz2000}. Their analysis results
\footnote{We notice that the star labeled 1 in ELC is the one which
is closest to the observer at primary (deepest) eclipse, usually at phase zero, and thus star 2 is eclipsed at that
time. This labeling scheme is opposed to that used traditionally in light curves of eclipsing binaries, which we
adopted. Thus, star 1 in ELC is the secondary component, labeled as 2 in this paper.} in
$T_{\rm eff1}$ = 3920 K, $M_1$ = 0.61 $M_\sun$, $R_1$ = 0.67 $R_\sun$;
$T_{\rm eff2}$ = 4120 K, $M_2$ = 0.67 $M_\sun$, $R_2$ = 0.79 $R_\sun$.

\cite{wolf2010} analyzed new BVR light curves using PHOEBE \citep[PHysics Of Eclipsing BinariEs,][]{prsa2005} obtaining 
$T_{\rm eff1}$ = 3920 K (adopted), $M_1$ = 0.61 $M_\sun$ (adopted), $R_1$ = 0.71 $R_\sun$;
$T_{\rm eff2}$ = 3825 K, $M_2$ = 0.67 $M_\sun$ (adopted), $R_2$ = 0.67 $R_\sun$.
These latter authors adopted $T_{\rm eff1}$, $M_1$, and $M_2$ from the \cite{coughlin2007} solution.
More recently, \cite{zhang2014} analyzed their own VRI light curves but they did not provide absolute masses
and radii values in their paper. One year later, \cite{zhang2015} also observed this system and claimed the detection
of a variation in the orbital period. All these previous studies analyzed only LCs and they did not publish RV orbital solutions.

In this paper we present the first complete physical measurement of the \nsvs\ system parameters.
We used optical VRI band and infrared (IR) JK band LCs jointly with  RVs and 
photometric measurements to obtain out-of-eclipse reliable photometric color indices of the system.
This paper is organized as follows: In Sect.~\ref{sec:LCobs} the LC data acquisition is described.
Section~\ref{sec:RVobs} is devoted to RV data acquisition. Section~\ref{sec:analysis} is devoted
to the analysis of the stellar observables and the RV and LC fits, while Sect.~\ref{sec:system}
describes the main results and the physical parameters obtained.
Throughout this paper, we use the subscript 1 for the primary star, that is, the one eclipsed in the photometric primary
(deepest) minimum, and the subscript 2 for the secondary star.

\section{Light-curve observations}
\label{sec:LCobs}

\subsection{Optical differential photometry}
\label{subsec:Opticaldphot}

\nsvs\ was observed by \cite{coughlin2007} using the Southeastern Association for Research
in Astronomy (SARA) 0.9 m telescope at Kitt Peak National Observatory (USA) in Johnson $V$, $R,$ and $I$
filters. Given the high quality of those observations they are suitable to properly model this
system. This $VRI$ differential photometry (given in Table~\ref{table:SARAphot}) includes two primary and four
secondary minima.

\onltab{
\begin{table*}
\caption{SARA $VRI$ bands differential photometry for \nsvs\ . The full version of this table is available in
electronic form at the CDS.}
\label{table:SARAphot}
\centering
\begin{tabular}{cccc}
\hline  \hline
HJD             &$\Delta$m  &$\sigma$ &Filter\\
(day)          &(mag)    &(mag)   &--\\
\hline
2453856.690192 &$-$0.842 &0.004 &I\\
2453856.690875 &$-$0.417 &0.004 &R\\
2453856.692194 &0.015    &0.004 &V\\
2453856.695030 &$-$0.845 &0.004 &I\\
2453856.695944 &$-$0.411 &0.004 &R\\
2453856.697784 &0.021    &0.004 &V\\
...&\\
\hline
\end{tabular}
\end{table*}
}

\subsection{Infrared differential photometry}
\label{subsec:IRdphot}

Infrared photometry was carried out using the CAIN (CAmara INfrarroja) near-IR (NIR) camera placed at the Cassegrain focus
of the 1.5~m Carlos S\'anchez Telescope (TCS) at Teide Observatory, Canary Islands \citep{cabrera-lavers2006}.
This instrument is built around a 256x256 element NICMOS 3 HgCdTe array, with 40 $\mu m$ square pixels.
The instrument is cooled to the  temperature of liquid nitrogen ($LN_2$) and can operate in two optical configurations:
\textit{wide} and \textit{narrow}. We chose to operate the \textit{wide} optics, which provides a field of
4.25$\times$4.25 arcmin, in order to obtain a field with enough comparison stars. With this optical configuration,
the plate scale is 1.0 $arcsec/pix$. The filter wheel is equipped with $JHK$ filters, among other NIR filters.
These filter passbands are close to those defined by the Johnson photometric system \citep{johnson1966,glass1985}.
The readout electronics can operate in several modes but we chose to operate in the standard Fowler (8,2) mode.
In order to subtract the sky background, we used a dithering pattern of four pointings using the guiding system of
the telescope.

At the beginning of each night bright and dark dome flat fields were acquired to correct for variations in the pixel sensitivity.
Bright flat fields were taken in the standard way against the inner part of the dome with the lights switched on.
Dark flat fields were taken immediately with the same configuration but with the lights switched off.
This is done to avoid changes in ambient temperature, and is equivalent to observing dark frames in the optical spectrum.

The images were processed using standard IR reduction techniques which include bright and dark flat-field processing,
bad pixel masking, science image registration, background subtraction, and aperture photometry.
Because of the plate scale of the instrument and the great number of bad pixels present in the array, bad-pixel correction is particularly important. For this task we created a custom bad-pixel map by
using the ratio of long and short dome flat fields.
We automated all these steps within a custom IRAF reduction pipeline devoted to the reduction and aperture
photometry of large sets of CAIN images.

For each reduced image we performed aperture photometry for a set of aperture radii centered at each object and
selected the one providing the best signal-to-noise ratio (S/N), taking into account the full width at half
maximum (FWHM) of each image and the electronic parameters of the camera.
This approach has the advantage of extracting the best photometry with varying seeing and atmospheric conditions
but is more computationally intensive. The S/N of the differential photometry is limited
in this case by the absence of suitable comparison stars of similar brightness to the target in the small
field of the CAIN camera at the J and K band wavelengths. The results of this photometry are given
in Table~\ref{table:TCSphot}.

\onltab{
\begin{table*}
\caption{TCS $JK$ bands differential photometry for \nsvs. The full version of this table is available in
electronic form at the CDS.}
\label{table:TCSphot}
\centering
\begin{tabular}{ccccc}
\hline  \hline
HJD             &$\Delta$m      &$\sigma$&Airmass&Filter\\
(day)           &(mag)          &(mag)  &--     &--\\
\hline
2453846.559472  &$-$2.332       &0.002  &1.178  &J\\
2453846.560155  &$-$2.252       &0.002  &1.175  &J\\
2453846.560884  &$-$2.297       &0.002  &1.173  &J\\
2453846.561532  &$-$2.270       &0.002  &1.171  &J\\
2453846.562478  &$-$2.609       &0.001  &1.167  &K\\
2453846.563450  &$-$2.645       &0.001  &1.164  &K\\
2453846.564434  &$-$2.443       &0.001  &1.160  &K\\
2453846.565384  &$-$2.569       &0.001  &1.157  &K\\
...&\\
\hline
\end{tabular}
\end{table*}
}

\section{Radial-velocity observations}
\label{sec:RVobs}

\nsvs\ was placed on the observing program at the Harvard-Smithsonian
Center for Astrophysics in June of 2009 and was monitored
spectroscopically until May of 2013 with the Tillinghast Reflector
Echelle Spectrograph \citep[TRES;][]{furesz2008}, a bench-mounted
fiber-fed echelle spectrograph attached to the 1.5 m Tillinghast
reflector at the Fred L.\ Whipple Observatory on Mount Hopkins
(Arizona, USA). This instrument delivers a wavelength coverage of
$\sim$3800--9100~\AA\ in 51 orders, at a resolving power of $R \approx
44,000$. A total of 38 spectra were obtained with S/Ns ranging from about 11 to 45 per resolution element of
6.8~\kms\ . The observations were reduced with a custom pipeline
\citep[see][]{buchhave2010}, and the wavelength calibration was
carried out based on exposures of a thorium-argon lamp before and
after each science frame.

Radial velocities for the two components of \nsvs\ were computed with
the 2D cross-correlation algorithm TODCOR
\citep{zucker1994}, using templates selected from a large library of
synthetic spectra based on models by R.\ L.\ Kurucz
\citep[see][]{nordstrom1994, latham2002}. For these determinations
we used only the echelle order centered at $\sim$5187~\AA\ 
(containing the \ion{Mg}{1}~b triplet), given that previous experience shows it contains
most of the information on velocity, and because our template library
is restricted to this wavelength range.  The optimal template for each
star was found by running grids of cross-correlations over a wide
range of effective temperatures ($T_{\rm eff}$) and rotational
broadening ($v \sin i$) following \cite{torres2002b}, and selecting
the combination giving the highest cross-correlation value averaged
over all observations and weighted by the strength of each
exposure. In this way we estimated temperatures of $4600 \pm 150$~K
for both stars, and projected rotational velocities of $67 \pm
3$~\kms\ and $66 \pm 3$~\kms\ for the primary (star eclipsed at the
deeper minimum) and secondary. We adopted $\log g$ values of 4.5 for
both stars, and solar metallicity. The resulting RVs in
the heliocentric frame and corresponding uncertainties are listed in
Table~\ref{table:RV}.

\onltab{
\begin{table*}
\caption{Heliocentric RV measurements of \nsvs. The full version of this table is available in
electronic form at the CDS.}
\label{table:RV}
\centering
\begin{tabular}{cccccc}
\hline
\hline
HJD               &$RV_1$       &$\sigma_1$     &$RV_2$         &$\sigma_2$         &Orbital\\
(day)             &(\kms)       &(\kms)         &(\kms)         &(\kms)          &phase\\
\hline
2454993.8051 &    111.32        &  3.35 &       $-$143.88       &  3.07         &  0.8210 \\
2454995.7535 &    $-$148.72     &  2.72 &       115.13          &  2.49         &  0.2958 \\
2455243.0366 &    $-$150.97     &  3.61 &       108.29          &  3.30         &  0.3042 \\
...&\\
\hline
\end{tabular}

\end{table*}
}

\section{Analysis of the system}
\label{sec:analysis}

\subsection{Distance and reddening}
\label{subsec:reddening}

\begin{table}
\caption{Computed reddening values for \nsvs.}             
\label{table:reddening}      
\centering                          
\begin{tabular}{c c c c}        
\hline\hline                 
$E_{\infty}(B-V)$       &$E_d(B-V)$     & Reference\\    
                        &(d=131.6 pc)   &\\
\hline                        
0.0405$\pm$0.0007       &0.0204$\pm$0.0004      &SF2011\tablefootmark{a}\\      
0.0470$\pm$0.0008       &0.0237$\pm$0.0004      &SFD1998\tablefootmark{b}\\
0.050$\pm$0.020         &0.025$\pm$0.010        &G2018\tablefootmark{c}\\
\hline
Mean value (adopted)    &$0.0231\pm0.0025$      &\\
\hline                                   
\end{tabular}
\tablefoot{\\
\tablefoottext{a}{\cite{schlafly2011}}\\
\tablefoottext{b}{\cite{schlegel1998}}\\
\tablefoottext{c}{\cite{green2018}}
}
\end{table}

The Gaia parallax for \nsvs\ is $\pi=7.571\pm0.043$ milli-arcseconds (mas) which translates
to a distance of $d=131.6\pm0.8$ pc \citep{bailer-jones2018}. We used recent Galactic
reddening maps \citep{schlafly2011, schlegel1998, green2018} to compute a mean reddening
of $E(B-V)=0.0231\pm0.0025$ (see Table~\ref{table:reddening}), scaled for the Gaia distance using the equation
\citep{bilir2008, bahcall1980}:

\begin{equation}\label{eq:reddening_d}
E_d(B-V)=E_{\infty}(B-V)\left[1-\exp\left(-\frac{|d\sin b|}{h}\right)\right]
.\end{equation}

In this equation, $E_d(B-V)$ is the reddening at the distance $d$, $E_{\infty}(B-V)$ is the total
reddening along the line of sight obtained from the reddening map, $d$ is the Gaia distance to \nsvs,
$b$ is the Galactic latitude, and $h$ is the Galactic scale height, taken as $h=$125 pc.
Equation~\ref{eq:reddening_d} is obtained from a model composed from a disk with decreasing density with the
distance to the Galactic plane, that is, it assumes a smooth distribution of the interstellar absorption along
the line of sight between the observer and the eclipsing binary, which may not be the case.
For this reason, the actual uncertainty in $E_d(B-V)$, taken as
the mean of the three values, may be larger than the simple mean adopted here.

\subsection{Effective temperature}
\label{sec:Teff}

As in the case of other faint LMDEBs \citep[see, e.g., ][]{iglesias-marzoa2017} we gathered photometry
from several catalogs for \nsvs\ and found discrepancies which prevent us from using the optical
data from those catalogs to obtain reliable color indices, and subsequent $T_{\rm eff}$ estimates.
Specifically, there are inconsistencies among optical $BVR$ magnitudes from different catalogs
and most of them do not provide the time of the measurements to check if the system was undergoing
eclipses at those times.
To deal with these problems, we performed standardized photometric measurements to obtain consistent color
indices in the optical $BVR_CI_C$ bands for a set of LMDEBs. The observations were done using the full
frame mode of the Tr\"omso CCD Photometer \citep[TCP, ][]{ostensen2000thesis,ostensen2000} 
at the IAC80 telescope (Teide Observatory, Canary Islands, Spain) during two photometric nights.
Full details of these measurements, including the transformation equations and coefficients, were
published in \cite[][Sect.4.1]{iglesias-marzoa2017}. \nsvs\ was observed the night of June 26 2012 at
photometric phases in the range 0.881$-$0.895, well in out-of-eclipse phase.
The resulting $BV R_C I_C$ magnitudes are shown in Table~\ref{table:calibrated_phot}.
In the same table we also list the observed color indices, including the NIR $JHK_S$
bands from Two Micron All Sky Survey \citep[2MASS, ][]{skrutskie2006}. The 2MASS photometry was obtained at JD 2450868.0374,
which corresponds to an orbital phase of 0.868, using the ephemeris of \cite{zhang2015} (see Sect.~\ref{subsec:ephem}).

\begin{table}
\caption{Calibrated observed magnitudes and color indices for \nsvs\ 
obtained from our absolute photometry measurements. The IR magnitudes
are taken from 2MASS.}
\label{table:calibrated_phot}      
\centering                          
\begin{tabular}{l c}        
\hline\hline                 
Band            &Adopted value (mag)\\    
\hline                        
$B$             &14.008$\pm$0.007\\
$V$             &12.843$\pm$0.005\\
$R_C$           &12.088$\pm$0.006\\
$I_C$           &11.379$\pm$0.008\\
$J$\tablefootmark{a}            &10.326$\pm$0.021\\
$H$\tablefootmark{a}            &9.707$\pm$0.019\\
$K_S$\tablefootmark{a}          &9.547$\pm$0.017\\
$B-V$           &1.165$\pm$0.009\\
$V-R_C$         &0.755$\pm$0.008\\
$V-I_C$         &1.464$\pm$0.009\\
$V-K_S$         &3.296$\pm$0.018\\
$R_C-I_C$       &0.709$\pm$0.010\\
$I_C-K_S$       &1.832$\pm$0.019\\
$J-H$           &0.619$\pm$0.028\\
$J-K_S$         &0.779$\pm$0.027\\
$H-K_S$         &0.160$\pm$0.025\\
\hline                                   
\end{tabular}
\tablefoot{\\
\tablefoottext{a}{Identifier: 2MASS J16072787+1213590.}
}
\end{table}

Using the adopted mean value of $E(B-V)$, we computed the interstellar extinction in all bands using
Table~6 of \cite{schlafly2011}, and dereddened all the color indices to obtain the values shown in
Table~\ref{table:Teff_calibrations}.
We used the empirical calibrations of \cite{casagrande2010} and \cite{huang2015} to obtain the $T_{\rm eff}$ 
values for each color index, adopting $[Fe/H]=0.0$ for both calibrations.
A constant offset of 130 K was subtracted from the $T_{\rm eff}$ values obtained from
\cite{casagrande2010}, as noted by \cite{huang2015} (see their Sect.~4.1). The computations for the
\cite{huang2015} calibration were done using the FGKM dwarfs coefficients.

\begin{table}
\caption{Mean effective temperature estimations resulting from our
$BVR_CI_C$ photometry, and 2MASS photometry for \nsvs.
The color indices have been dereddened from those in Table~\ref{table:calibrated_phot} (see text).}
\label{table:Teff_calibrations}
\centering
\begin{tabular}{lll}
\hline
\hline
Index           &Index value            &$T_{\rm eff}$ (K)\\
\hline
\multicolumn{3}{c}{Huang et al 2015}\\
\hline
$B-V$           &1.145$\pm$0.015        &4257$\pm$77\\
$V-R_C$         &0.742$\pm$0.012        &4256$\pm$106\\
$V-I_C$         &1.436$\pm$0.012        &4259$\pm$103\\
$V-J$           &2.470$\pm$0.023        &4126$\pm$88\\
$V-H$           &3.083$\pm$0.021        &4133$\pm$82\\
$V-K_S$         &3.240$\pm$0.019        &4141$\pm$79\\
$R_C-I_C$       &0.694$\pm$0.012        &4264$\pm$119\\
\hline
\multicolumn{3}{c}{Casagrande et al 2010\tablefootmark{a}}\\
\hline
$B-V$           &1.145$\pm$0.015        &4413$\pm$79\\
$V-R_C$         &0.742$\pm$0.012        &4303$\pm$66\\
$V-I_C$         &1.436$\pm$0.012        &4189$\pm$60\\
$J-K_S$         &0.770$\pm$0.027        &4201$\pm$154\\
\hline
Mean value (adopted)    &               &4240$\pm$100\\
\hline
\end{tabular}
\tablefoot{\\
\tablefoottext{a}{All $T_{\rm eff}$ values with a constant offset of 130 K subtracted.}
}
\end{table}

The resulting mean $T_{\rm eff}$ value of 4240$\pm$100 K in Table~\ref{table:Teff_calibrations} is in line with the Gaia value
of 4316 K but is significantly cooler than the spectroscopic value (4600$\pm$150 K). There are several possible explanations for this discrepancy: a third light caused by a cool object contaminating the photometry, a significantly low metallicity,
or a RV template mismatch. The first possibility was discarded by the third light tests shown in Sect.~\ref{subsec:thirdlight}.
In the analysis of the RV we assumed solar metallicity but a lower value of $[Fe/H]=-0.5$ dex would yield a spectroscopic $T_{\rm eff}$
about 300 K cooler than that obtained. Unfortunately, there is not spectroscopic $[Fe/H]$ estimations for
this system, and they would  be difficult to measure for this object because of the rotationally broadened lines.
The space motion of the system (see Sect.~\ref{sec:system}) places it in the Galactic thin disk but this position
does not rule out a low-metallicity system. Indeed, the photometric metallicity computed using
the $J-K$ and $V-K$ indices and the relation of \cite{mann2013} points towards a low metallicity
of $[Fe/H]=-0.31\pm0.21$ dex, though it must be pointed out that the computation is near the validity
limit of that calibration. Finally, the synthetic spectra templates used to perform the cross-correlation start to
differ from real stars below $T_{\rm eff}\lesssim4300-4500$ K due to the presence of an
increasing number of spectral lines. A combination of these two effects (low metallicity,
and template mismatch) may be the reason for the reported $T_{\rm eff}$ difference,
which in any case does not  affect the velocities significantly.

\subsection{Period and ephemeris}
\label{subsec:ephem}

We selected the linear ephemeris of \cite{zhang2015} to phase our light curves:

\begin{equation}\label{eq:lin_ephem}
MinI(HJD)=2453274.1705(5)+0.5607222(2)\times E
.\end{equation}

This ephemeris equation was computed using the available minima published in the literature,
including the observed eclipses from \cite{coughlin2007} used in the LC analysis in Sect.~\ref{sec:analysis},
and the IR photometry.

In their analysis, \cite{zhang2015} found a small decrement of the period at a rate of
$dP/dt=-2.79\times10^{-7}$ d/yr, which was not included in our model because of the small time span of the
LC observations used. In order to confirm this behavior, we searched for new minima in a number of photometric surveys. We found observations of this system in the Palomar Transient Factory \cite[PTF, ][]{law2009PASP},
the Catalina Real Time Transient Survey \cite[CRTS, ][]{drake2009ApJ},
and the All-Sky Automated Survey for Supernovae \citep[ASAS-SN, ][]{kochanek2017PASP}. Unfortunately,
these surveys have low time resolution, and the measurements are too sparse to get accurate mid-eclipse times.
Luckily, we found new CCD minima published by \cite{honkova2015OEJV}, \cite{jurysek2017OEJV}, and
\cite{smelcer2019BRNOdatabase} in the BRNO database\footnote{\url{http://var2.astro.cz/EN/brno/index.php?lang=en}}.

\begin{figure}[t]
  \centering
\includegraphics[width=\hsize]{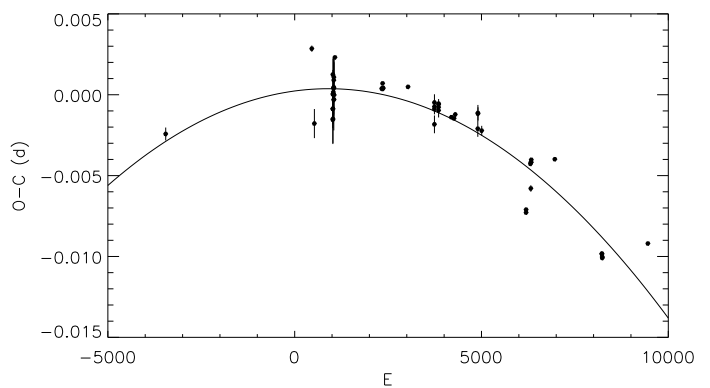}
\caption{O-C diagram with all the points collected from literature (see table
\ref{table:ecl_times}) and the resulting quadratic fit (continuous line).}
  \label{fig:ocfit}
\end{figure}

In Fig.~\ref{fig:ocfit} we show the $O-C$ diagram computed from the ephemeris in Eq.~\ref{eq:lin_ephem} for all of
the minima in the literature, both primary (P) and secondary (S) eclipses, and our IR light curves.
In our case, the times were obtained using the Kwee \& van Woerden method \citep{kwee1956BAN}.
All the observed eclipse times are also listed in Table~\ref{table:ecl_times}. The points
published in the BRNO database confirm the parabolic trend and therefore also the period change.
A quadratic fit to all the points is also shown in Fig.~\ref{fig:ocfit} and updates the quadratic ephemeris
published by \cite{zhang2015}, resulting in primary minimum times given by the equation:

\begin{eqnarray}\nonumber
\label{eq:quad_ephem}
MinI(HJD)=2453274.17073(58)+0.56072251(25) \times E \\
-(1.72\pm0.18)\times10^{-10} \times E^2
.\end{eqnarray}

The new period change, obtained from the quadratic coefficient, is $dP/dt=-(2.23\pm0.24)\times10^{-7}$ d/yr,
which is a typical value seen in other LMDEBs \citep[see for example][]{lee2013AJ}.
Since this is a well-detached binary system (see Sect.~\ref{sec:analysis}) this period change cannot
be due to mass exchange between components.

\nsvs\ is composed of two active stars, and therefore a possible explanation for its period change
could be angular momentum loss (AML) due to magnetic braking \citep{bradstreet1994ASPC}.
We computed the rate of period change produced by AML using Eq. 2 of \cite{bradstreet1994ASPC}
taking $k^2=0.1$ and the results in Table~\ref{table:AbsDimensions}.
This resulted in $(dP/dt)_{AML}=-1.1\times10^{-10}$ d/yr, a value too low to
account for the observed variation. Another possible explanation is the presence of a third body
in the system in a long-period orbit.
This would have to be a low-luminosity object, since no measurable third light is
detected in this system (see Sect.~\ref{subsec:thirdlight}).
The third body scenario is also supported by the observation of regular period changes
in other LMDEB attributed to substellar companions \citep{wolf2016AA}.
However, to confirm this possibility a periodic behavior in the $O-C$ diagram is required,
and this is not seen in this eclipsing binary.

\onltab{
\begin{table*}
\caption{\nsvs\ eclipse times from the literature and from our IR photometry.
In the third column P stands for primary eclipse and S for secondary eclipse.
The full version of this table is available in electronic form at the CDS.}
\label{table:ecl_times}
\centering
\begin{tabular}{cccc}
\hline
HJD             &$\sigma$       &Eclipse type &Reference\\
(d)             &(d)            &-              &-\\
\hline
\hline
2451338.8354    &0.0004         &S      &1\\
2453570.5104    &0.0009         &S      &2\\
2454581.49469   &0.00016        &S      &3\\
2454592.42875   &0.00009        &P      &3\\
2454594.39162   &0.00009        &S      &3\\
...&\\
\hline
\end{tabular}
\tablebib{(1)~\citet[NSV, ][]{kazarovets1998IBVS};
(2) \citet[ASAS, ][]{pojmanski1998AcA};
(3) \citet{wolf2010};
(4) \citet{coughlin2007};
(5) \citet{zhang2014};
(6) \citet{zhang2015};
(7) \citet{honkova2015OEJV};
(8) \citet{jurysek2017OEJV};
(9) \citet{smelcer2019BRNOdatabase};
(10) this work, from IR photometry.}
\tablefoot{In reference 7,  the minimum at 2456824.37893 appears as P but it must be S. It is corrected in this table.}
\end{table*}
}

\subsection{Rotation and synchronicity parameter}
\label{subsec:rotation}

We assumed that the two components of this system are tidally synchronized \citep{hut1981}.
This is a reasonable assumption, as the synchronicity process is usually faster than the
circularization process for convective envelope stars, and the secondary eclipse timings for
the system of \nsvs\ suggest that the orbit is already circularized.
As a result, the synchronicity parameters for this system were
set to $F_1=F_2=1$. This assumption is verified in Sect.~\ref{sec:system}.

\subsection{Third light}
\label{subsec:thirdlight}

Given the discrepancy between the optical and spectroscopic effective temperatures and the reported
detection of a period variation, we searched extensively for a third light in this system.
The existence of a third light could bias the mean color of the binary
system to lower temperatures and is correlated with the inclination, since the eclipse depths of
a system with third light can be reproduced with a similar system model but at lower inclination. 
All our tests, fitting for different $T_{\rm eff}$ in the components, 
were also repeated fitting for third light with the other parameters, always with negative results.
We also fitted the $VR_CI_C$ LCs from \cite{zhang2014} with and without third light, but the solutions
were in all cases compatible with no third light.

\subsection{Radial-velocity fit}
\label{subsec:RV_fit}

The RV observations were fitted using the \verb|rvfit| code, which can simultaneously
fit the seven parameters of a double-line spectroscopic binary using
an adaptive simulated annealing algorithm \citep[see][for details]{iglesias-marzoa2015}.
For the case of \nsvs\ we adopted a circular orbit model, given that the secondary
eclipse lies on phase 0.5. Thus, we fixed the eccentricity to zero and the argument of
the periastron, which is undefined for a circular orbit, to $\omega$=90 degrees to match
the times of the eclipses.

\begin{figure}
  \centering
\includegraphics[width=\hsize]{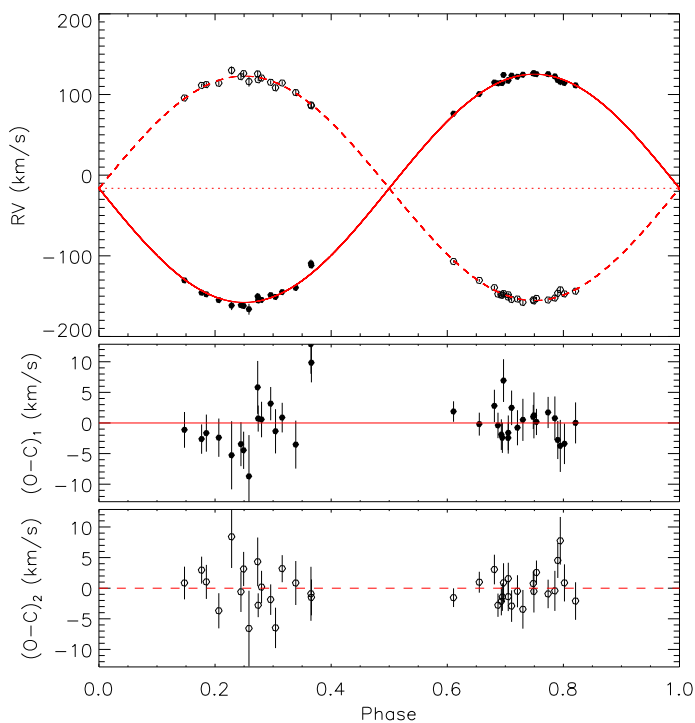}
\caption{Spectroscopic orbit (top panel) and residuals (bottom panel)
resulting from the fit to the RVs observed by TRES.
The eccentricity was fixed to zero according to a circular orbit.
The resulting parameters are those of Table~\ref{table:RVresults}.
The observations are plotted with their uncertainties, though in the
top panel they have nearly the same size as the symbols.}
  \label{fig:RVfit}
\end{figure}

Given that the orbital period is known from photometry, we fitted the remaining three
parameters ($\gamma$, $K_1$, $K_2$) to the observations. Once the solution was found,
we computed robust uncertainties in these three parameters using a Markov Chain Monte-Carlo (MCMC)
procedure. The fitted RV orbital solution is shown in Fig.~\ref{fig:RVfit} and the
obtained parameters and derived physical quantities in Table~\ref{table:RVresults}.
The value obtained for the mass ratio $q=M_2/M_1$ shows that the secondary component,
that is, the one eclipsed at the photometric secondary minimum, is slightly more massive than the primary,
despite having less surface luminosity.

\begin{table}
\caption{Radial-velocity fitting results.}
\label{table:RVresults}      
\centering                          
\begin{tabular}{l r}        
\hline\hline                 
Parameter                       &Value\\
\hline
\multicolumn{2}{c}{Adjusted Quantities}\\
\hline
$P$ (d)                         &0.56072200\tablefootmark{a}\\
$T_p$ (HJD)                     &2454933.90821\tablefootmark{a}\\
$e$                             &0.0\tablefootmark{a}\\
$\omega$ (deg)                  &90.0\tablefootmark{a}\\
$\gamma$ (km/s)                 &$-$16.31 $\pm$ 0.31\\
$K_1$ (km/s)                    &141.46 $\pm$ 0.48\\
$K_2$ (km/s)                    &139.09 $\pm$ 0.49\\
\hline
\multicolumn{2}{c}{Derived quantities}\\
\hline
$M_1\sin ^3i$ ($M_\sun$)        &0.6360 $\pm$ 0.0050\\
$M_2\sin ^3i$ ($M_\sun$)        &0.6469 $\pm$ 0.0050\\
$q = M_2/M_1$                   &1.0170 $\pm$ 0.0050\\
$a_1\sin i$ ($10^6$ km)         &1.0907 $\pm$ 0.0037\\
$a_2\sin i$ ($10^6$ km)         &1.0725 $\pm$ 0.0038\\
$a  \sin i$ ($10^6$ km)         &2.1632 $\pm$ 0.0053\\
\hline
\multicolumn{2}{c}{Other quantities}\\
\hline
$\chi^2$                        &84.35\\
$N_{obs}$ (primary)             &38\\
$N_{obs}$ (secondary)           &38\\
Time span (days)                &1449.96\\
$rms_1$ (km/s)                  &3.97\\
$rms_2$ (km/s)                  &3.17\\
\hline
\end{tabular}
\tablefoot{
\tablefoottext{a}{Parameter fixed beforehand.}
}
\end{table}

The two deviating RV observations in the primary star RV curve near phase 0.36 caught
our attention. We checked these spectra looking for abnormal features or for contamination from the
Moon but nothing appeared to be out of the ordinary so we decided to keep them for the fit.
The contamination from moonlight can bias the RVs because the extra lines from
the reflected light from the Sun can be blended with the lines of the binary components
to different degrees, thus distorting the line profiles and therefore affecting
the determination of the centroids in the cross-correlation analysis.
As a check, we repeated the fit without these two points and obtained values of 
$\gamma=-16.52\pm0.31$ \kms, $K_1=141.78\pm0.49$ \kms, and $K_2=139.01\pm0.49$ \kms, consistent
within $1\sigma$ with the adopted solution. The resulting mass ratio from this
solution is $q=1.0199\pm0.0050$, confirming that the secondary is more massive than the primary as
before, and that the two points are not the cause of the $q>1$ value.

Previous estimates of the $q$ parameter reported in the literature give very different values,
as they were done fitting only LCs. For detached eclipsing binaries the value of $q$ is not constrained by the LCs,
and it could happen that the value that numerically minimizes the merit function for the LCs does not correspond
to the actual $q$ value of the system.
For example \cite{zhang2015} did a search of the $q$ value in the range 0.2-1.0, which resulted in
$q=0.280\pm0.004$. The closest value of $q$ found in the literature was obtained
by \cite{coughlin2007} who reported $q=$0.91.

\subsection{Light-curve fit}
\label{subsec:LC_fit}

\begin{figure}
\centering
\includegraphics[width=\hsize]{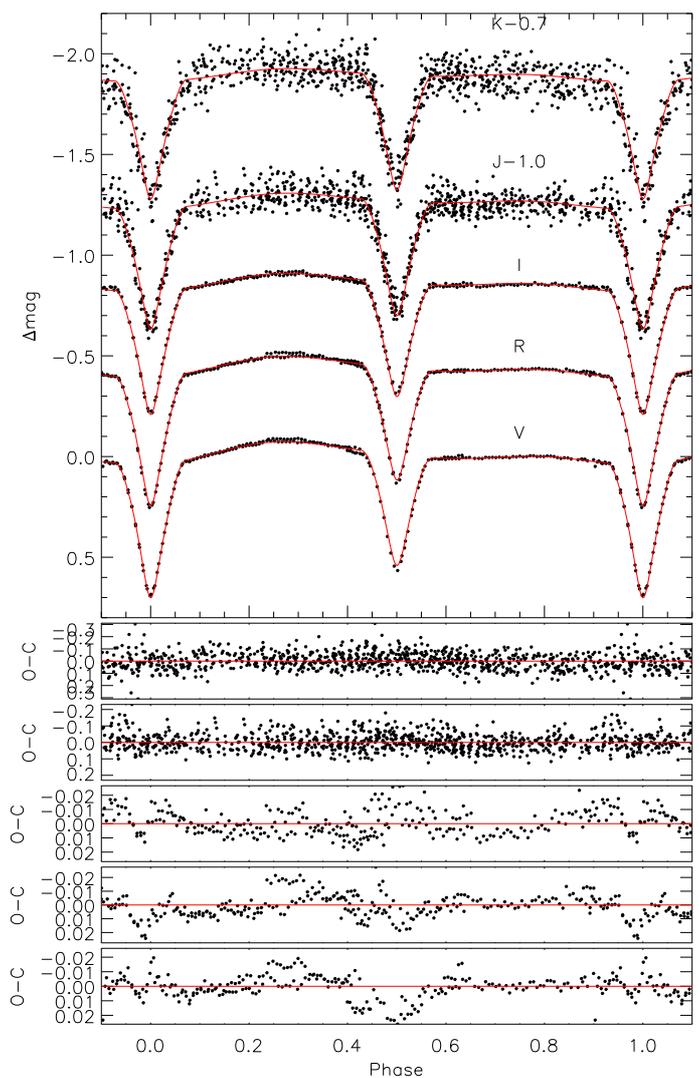}
  \caption{Top panel: \nsvs\ light curves (points) and PHOEBE fitted model (red lines)
  with the parameters of Table~\ref{table:LCresults}.
  From top to bottom: $K$, $J$, $I$, $R$, and $V$ differential light curves. $K$ and $J$ band TCS filters are displaced
  -0.7 and -0.1 magnitudes, respectively, for a better viewing. Lower panels: Residuals of the fits
  in the same order as the light curves in the top panel. We note the different vertical scales of the panels.}
  \label{fig:cfase_modelos}
\end{figure}

{\renewcommand{\arraystretch}{1.2}
\begin{table*}
\caption{\nsvs\ parameters computed from the PHOEBE fit to the VRIJK differential light curves.}
\label{table:LCresults}
\begin{center}
\begin{tabular}{lcc|lcc}
\hline
\hline
Parameter                       &Primary                &Secondary                              &Parameter                      &Primary                &Secondary\\
\hline
\multicolumn{3}{c|}{{\bf Geometric and orbital parameters}}                                     &$x_1$,$y_1$ ($R$ band)              &0.457, 0.330           &0.351, 0.450\\
$P$ (d)                         &\multicolumn{2}{c|}{0.5607222\tablefootmark{a}}                &$x_1$,$y_1$ ($I$ band)              &0.249, 0.455           &0.157, 0.555\\
$T_0$ (HJD)                     &\multicolumn{2}{c|}{2453274.1705\tablefootmark{a}}             &$x_1$,$y_1$ ($J$ band)              &0.094, 0.544           &0.037, 0.586\\
$\Delta\phi$                    &\multicolumn{2}{c|}{0.000219$\pm$0.000051}                     &$x_1$,$y_1$ ($K$ band)              &$-$0.156, 0.711        &$-$0.147, 0.673\\
$q$                             &\multicolumn{2}{c|}{1.0170\tablefootmark{a}}                   &\multicolumn{3}{c}{{\bf Spot 1 parameters (primary star)}}\\
$\gamma$ (\kms)                 &\multicolumn{2}{c|}{$-$16.31\tablefootmark{a}}                 &Colatitude (deg)           &31.4$\pm$3.3           &-\\
$i$ (deg)                       &\multicolumn{2}{c|}{86.465$\pm$0.083}                          &Longitude (deg)           &101.014$\pm$0.040      &-\\
$e$                             &\multicolumn{2}{c|}{0.0\tablefootmark{a}}                      &Radius (deg)                   &37.0$\pm$2.0           &-\\
$a$ ($R_\sun$)                  &\multicolumn{2}{c|}{3.1139$\pm$0.0076}                         &$T_{spot}/T_{surf}$            &0.884$\pm$0.020        &-\\
$\omega$ (deg)                  &\multicolumn{2}{c|}{90.0\tablefootmark{a}}                     &\multicolumn{3}{c}{{\bf Spot 2 parameters (primary star)}}\\
$\Omega$                        &5.559$\pm$0.021        &5.745$\pm$0.023                        &Colatitude (deg)           &57.2$\pm$2.0           &-\\
$F$                             &1.0\tablefootmark{a}   &1.0\tablefootmark{a}                   &Longitude (deg)           &355.145$\pm$0.044      &-\\
\multicolumn{3}{c|}{{\bf Radiative parameters}}                                                 &Radius (deg)                   &21.4$\pm$1.5           &-\\
$T_{\rm eff}$ (K)               &4240\tablefootmark{a}  &4120$\pm$100                   &$T_{spot}/T_{surf}$            &0.703$\pm$0.039        &-\\
$Albedo$                        &0.5\tablefootmark{a}   &0.5\tablefootmark{a}                   &\multicolumn{3}{c}{{\bf Spot 3 parameters (secondary star)}}\\
$\beta$                         &0.32\tablefootmark{a}  &0.32\tablefootmark{a}                  &Colatitude (deg)           &-                      &36.1$\pm$6.0\\
$l_3$ (for all bands)           &\multicolumn{2}{c|}{0.0\tablefootmark{a}}                      &Longitude (deg)           &-                      &0.0$\pm$2.4\\
\multicolumn{3}{c|}{{\bf Fractional radii}}                                                     &Radius (deg)                   &-                      &29.00$\pm$5.3\\
$r_{pole}$                      &0.2190$\pm$0.0010      &0.2130$\pm$0.0010                      &$T_{spot}/T_{surf}$            &-                      &0.940$\pm$0.021\\
$r_{point}$                     &0.2261$\pm$0.0011      &0.2190$\pm$0.0012                      &\multicolumn{3}{c}{{\bf Parameters computed from MCMC}}\\
$r_{side}$                      &0.2214$\pm$0.0010      &0.2151$\pm$0.0011                      &$\Delta\phi$                   &\multicolumn{2}{c}{0.00010$^{+0.00045}_{-0.00049}$}\\
$r_{back}$                      &0.2247$\pm$0.0011      &0.2179$\pm$0.0011                      &$i$ (deg)                   &\multicolumn{2}{c}{86.22$^{+0.61}_{-0.61}$}\\
$r_{vol}$                       &0.2218$\pm$0.0011      &0.2154$\pm$0.0014                      &$\Omega$                       &5.59$^{+0.12}_{-0.14}$ &5.71$^{+0.15}_{-0.12}$\\
\multicolumn{3}{c|}{{\bf Luminosities}}                                                         &$T_{\rm eff}$\tablefootmark{b} (K) &-                   &4104$^{+39}_{-55}$\\
$L/(L_1+L_2)\ (V)$              &0.5659$\pm$0.0022      &0.4341$\pm$0.0022                      &\multicolumn{3}{c}{{\bf Fractional volumetric radii from MCMC}}\\
$L/(L_1+L_2)\ (R)$              &0.5567$\pm$0.0023      &0.4433$\pm$0.0023                      &$r_{vol}$              &0.2205$^{+0.0054}_{-0.0076}$   &0.2156$^{+0.0059}_{-0.0071}$\\
$L/(L_1+L_2)\ (I)$              &0.5446$\pm$0.0024      &0.4554$\pm$0.0024                      &\multicolumn{3}{c}{{\bf Residuals from the fit}}\\
$L/(L_1+L_2)\ (J)$              &0.5371$\pm$0.0027      &0.4629$\pm$0.0027                      &$\sigma_V$ (mag)           &\multicolumn{2}{c}{0.008}\\
$L/(L_1+L_2)\ (K)$              &0.5305$\pm$0.0028      &0.4695$\pm$0.0028                      &$\sigma_R$ (mag)           &\multicolumn{2}{c}{0.008}\\
\multicolumn{3}{c|}{{\bf Limb-darkening coefficients (square-root law)}}                        &$\sigma_I$ (mag)           &\multicolumn{2}{c}{0.009}\\
$x_1$,$y_1$ (bol)               &0.227, 0.446           &0.140, 0.521                           &$\sigma_J$ (mag)           &\multicolumn{2}{c}{0.053}\\
$x_1$,$y_1$ ($V$ band)          &0.739, 0.067           &0.578, 0.250                           &$\sigma_K$ (mag)           &\multicolumn{2}{c}{0.069}\\
\hline
\end{tabular}
\end{center}
\tablefoot{
\tablefoottext{a}{Parameter fixed beforehand.}
\tablefoottext{b}{This uncertainty results from MCMC and doesn't take into account the one in the $T_{\rm eff1}$.}
}
\end{table*}
}

The $VRIJK$ light curves described in Sect.~\ref{subsec:Opticaldphot} and \ref{subsec:IRdphot}
were fitted simultaneously using the 1.0 SVN version of PHysics Of Eclipsing BinariEs \citep[PHOEBE,][]{prsa2005, prsa2011}.
PHOEBE is a front-end of the well-known Wilson-Devinney code \cite[WD,][]{wilson1971}
which adds improvements to the treatment of multiwavelengh light curves.
The WD employs a physical model of the gravitational distortions,
radiative properties, and spot parameters of an eclipsing binary star.
From preliminary fits we obtained fractional radii of $r_1=r_2\simeq0.22$,
and using $q=1.0170$ from RV curves, we estimated the Roche lobe radii for each component as
$r_{L1}\simeq0.35$ and $r_{L2}\simeq0.36$ using the \cite{eggleton1983} formula.
These values are well above the relative radii $r_1$ and $r_2$
and therefore we selected a ``detached eclipsing binary'' model.
To properly weight the fitting of each LC we measured the actual dispersion of each LC at quadratures
($\sigma_V=0.0047$, $\sigma_R=0.0047$, $\sigma_I=0.0058$, $\sigma_J=0.055$, $\sigma_K=0.063$)
and weighted each LC with its $\sigma$ value (curve-dependent weights).
This gives more weight in the model fit to the more precise VRI LCs.

Given the large number of parameters, we fixed some of them using external data and constraints.
The period and the epoch of the primary eclipse was fixed to the values of Eq.~\ref{eq:lin_ephem}.
The mass ratio $q=M_2/M_1$ and the system RV $\gamma$ were fixed to those of the RV fit
in Table~\ref{table:RVresults}. The eccentricity was fixed to $e=0.0$  corresponding to a circular orbit.
The synchronicity parameters for the two components were set to $F_{1,2}=1.0$  corresponding to tidally
synchronized stars \citep{hut1981}.

The $T_{\rm eff1}$ was set to the adopted value of the mean temperature of the system as computed
from the absolute photometry (4240$\pm$100 K, see Table~\ref{table:Teff_calibrations}). The albedos
were set to $A_1=A_2=0.5$ following the prescription for stars with convective envelopes \citep{rucinski1969b}.
The gravity-brightening $\beta_1$ and $\beta_2$ exponents were both fixed to 0.32
as suggested by \cite{lucy1967}. As a sanity check we computed the $\beta$ values following
the \cite{claret2000} study and obtained very similar values ($\beta_1=0.35$ for $T_{\rm eff1}=4240$ K
and $\beta_2=0.31$ for $T_{\rm eff1}=4120$ K from preliminary fits).

We selected a square-root limb-darkening (LD) law since this is the one that better fits
light curves at longer wavelengths and in the IR \citep{diaz-cordoves1992, vanhamme1993}.
The LD coefficients were automatically computed after each iteration using the tables of \cite{vanhamme1993}.
The third light was initially allowed to vary ($l_3\neq0$), but all our tests resulted in negative values
or values compatible with no third light, and so in the final fits it was fixed to zero
(see below). We also allowed for mutual heating effects between the components but they
do not significantly affect the fit.

\subsection{Spot modeling}
\label{subsec:spot_modeling}

The need to include spots in the LC model is evident in view of the modulation of the 
out-of-eclipse light curves. We made some preliminary tests placing spots in the two components at latitudes
of $\sim$45 degrees, given that there are hints that this is the range of latitudes most affected by spots
in low-mass stars \citep{hatzes1995IAUS,granzer2000}. The parameters of the spots, namely, latitude, longitude,
radius, and temperature ratio ($T_{spot}/T_{surf}$) were allowed to vary in order to fit the observed
variations of the light curves. In these initial tests we tried several spot scenarios,
including the use of bright spots (by forcing $T_{spot}/T_{surf}>1$)
facing the observer at phase $\sim$0.2 to reproduce the hump observed at that phase.
Finally, the best model was achieved with two dark spots ($T_{spot}/T_{surf}<1$) at convenient phases
to mimic the effect of the a bright spot at the opposite side of the binary. With these two dark spots,
the profile of the eclipses was much better reproduced, but small systematic residuals remained at
phases surrounding the secondary eclipse, and so we placed another spot in the secondary star which fitted
the residuals much better. Although PHOEBE can only fit two spots simultaneously,
we managed to fit the three spots by fitting two of them at once and cycling between them until
a satisfactory convergence was obtained. The model of \nsvs\ with spots in the two components 
is supported by the fact that the spectra show activity for the two components, since CaII
H and K lines are clearly seen in emission in both stars. Visual inspection shows
no difference in the height of the emission cores, which would suggest similar activity levels.
We fitted other models with spots in only one component but they did not fit as
well as the one with spots on the two stars.

\subsection{Final solution}
\label{subsec:final_solution}

We fitted the PHOEBE model allowing to vary the following parameters: the phase shift $\Delta \phi$,
the secondary effective temperature $T_{\rm eff2}$, the orbital inclination $i$,
the two surface potentials $\Omega_1$, $\Omega_2$, the passband luminosities (HLA),
and the spot parameters. The parameters obtained for the final solution are shown in
Table~\ref{table:LCresults}. The uncertainties in this table are the formal ones from the PHOEBE fit,
and they not include correlations among parameters or systematic effects.
The formal uncertainty in $T_{\rm eff2}$ is about 4 K, but taking into account that the
$T_{\rm eff1}$ uncertainty is $\pm$100 K, we choose to add them in quadrature to take into
account the dependence of the two values.

The $r_{vol}$ values are the volumetric radii from which the absolute radii can be computed.
We note that the stars are only slightly distorted ($r_{point}/r_{pole}\sim$3.2\% and 2.8\% for
the primary and the secondary, respectively) in spite of the proximity of the stars.
The fitted light curves and their residuals are shown in Fig.~\ref{fig:cfase_modelos}. 

\begin{figure}
\centering
\includegraphics[width=\hsize]{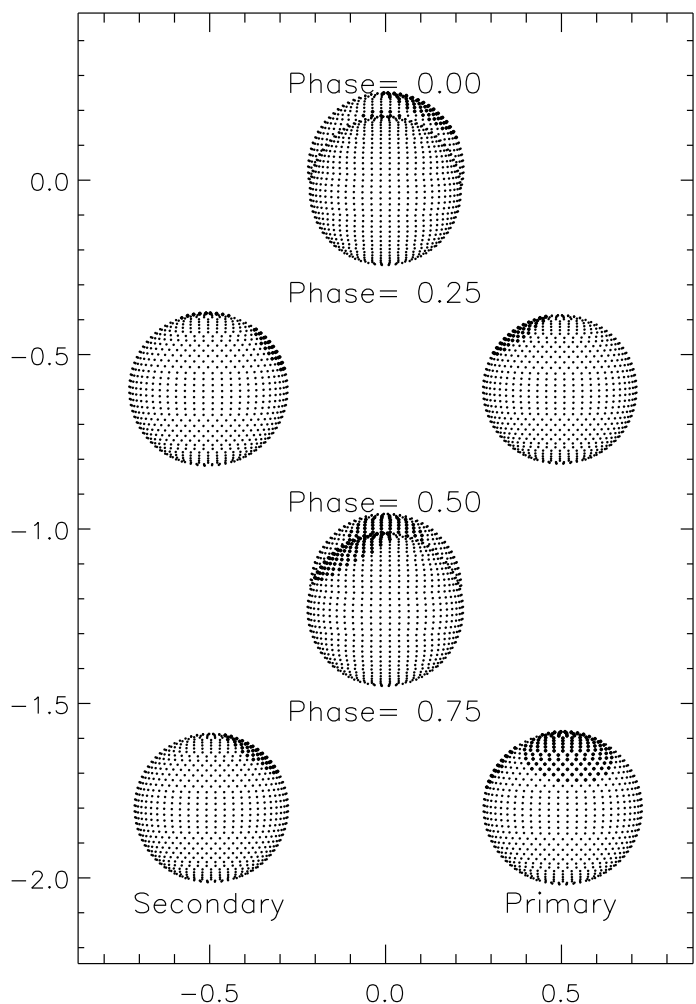}
  \caption{Representation of the spot configuration of \nsvs\ in the $(v,w)$ plane at four phases:
  from top to bottom, at the primary eclipse (phase 0.0), at the first quadrature (phase 0.25),
  at the secondary eclipse (phase 0.5), and at the second quadrature (phase 0.75).
  Both axes have units of relative radius and the images are displaced
  $-$0.60 from the previous one . The star in front is orbiting towards the right side and the
  primary star is the one with two spots.}
      \label{fig:mesh_model}
\end{figure}

Figure~\ref{fig:mesh_model} is a graphical representation of the spot configuration for four
orbital phases of \nsvs. It is possible that instead of extended spots (37, 21, and 29 degrees),
each of them comprises a group of close smaller spots with higher $T_{\rm eff}$ contrast
with the surrounding photosphere, or even that they are spots with variable surface $T_{\rm eff}$
distributions. However, with the present data, it is not possible to distinguish among these possibilities; this would require Doppler imaging observations \citep{strassmeier2009AARv}
as was done before in the case of YY Gem \citep{hatzes1995IAUS}.

A drawback of the PHOEBE code, which comes from the WD code, is that it does not allow  the radius ratio ($r_2/r_1$) to be fitted.
In partially eclipsing systems with similar components, as is the case here,
the radius ratio can often be poorly constrained and can be correlated with other parameters.
An analysis of this behavior is described in \cite{torres2002a} for YY Gem.
On the other hand, the sum of radii ($r_1+r_2$) is well constrained by the duration
of the eclipses, which can be measured with precision. Fitting separately for the individual radii
(or the potentials) is not optimal. Since the radius ratio is always strongly correlated with the light ratio,
it is often beneficial within MCMC to put a prior on the light ratio using the spectroscopic value,
which indirectly constrains the radius ratio. However, this has to be done
for the same mean wavelength of the spectra -- in this case 5187~\AA\ -- and cannot be done in PHOEBE because
the filter passbands do not match this wavelength.

Given that we assigned the mean effective temperature of the system to the primary star, the actual
effective temperature of the primary must be hotter than the imposed value. To check whether or not the assumed
$T_{\rm eff1}$ should be changed in a new analysis, we estimated the value of $T_{\rm eff1}$ using
the PHOEBE results and the equations for the bolometric luminosity, $L_1=4\pi R_1^2\sigma T_{\rm eff1}^4$,
and a similar equation for $L_2$.
The total bolometric luminosity of the system is $L_T=L_1+L_2=S_T \sigma T_{\rm eff m}^4$,
 $S_T$ being the total surface of the two components, $\sigma$ the Stefan-Boltzmann constant,
and $T_{\rm eff m}$ the mean effective temperature computed from the colors of the system.
From these equations we can compute the mean $T_{\rm eff m}$ as

\begin{equation}
\label{eq:Teff_decoupling}
T_{\rm eff m}=T_{\rm eff 1}\left[ \frac{1+L_2/L_1}{1+(L_2/L_1)(T_{\rm eff1}/T_{\rm eff2})^4} \right]^{1/4}
.\end{equation}

From the PHOEBE results in Table~\ref{table:LCresults} we computed the radii
$R_1=0.6907\pm0.0038\ R_{\sun}$, $R_2=0.6707\pm0.0047\ R_{\sun}$, and bolometric luminosities
$L_1=0.139\pm0.013\ L_{\sun}$, $L_2=0.117\pm0.012\ L_{\sun}$.
Using those values in Eq.~\ref{eq:Teff_decoupling} we obtained $T_{\rm eff1}=4300$ K for the primary component,
which is within $1\sigma$ of the value in the previous fit, and we fitted again the LCs
imposing this corrected $T_{\rm eff1}$.

The parameter values in Table~\ref{table:LCresults} result in a distance to the system of $d=136.1\pm6.9$ pc,
in very good agreement with the Gaia distance of $131.6\pm0.8$ pc (difference of $0.65\sigma$).
But from the parameters obtained from the new fit, fixing $T_{\rm eff1}=4300$ K, we estimated
a distance $d=147.2\pm6.8$ pc, which is $2.3\sigma$ greater than the Gaia distance.
In light of this result we prefer to maintain for the primary the initial $T_{\rm eff1}=4240\pm100$ K.
The difference in primary temperatures is well within the uncertainties in the mean photometric temperature.
This check also discards the value of 4600 K suggested by the spectra of the system
since the luminosity of a system with such $T_{\rm eff}$ would put the binary much further away than
is allowed by the constraint of the Gaia distance. All these tests were done fixing $l_3=0$, and we repeated them
by fitting for $l_3$ with the other parameters. Again, this resulted in no detectable third light.

A second test was done fitting for the $T_{\rm eff}$ of the two components exploiting the fact
that the LCs span a wide range of wavelengths. The constraint on the individual temperatures
are nevertheless too weak, and the fit resulted in two temperatures that are too low to account for the observed photometric colors.
As before, this fit was also repeated fitting for a third light without positive results.

We checked the individual $T_{\rm eff}$ of the two components using the relation among
fundamental parameters of \cite{mamajek2015} and the calibration of \cite{huang2015}.
We computed the absolute magnitudes for each component in several filters using the luminosity ratios
for each passband in Table~\ref{table:LCresults}, the out-of-eclipse calibrated magnitudes of
Table~\ref{table:calibrated_phot}, the Gaia distance for the system, and the computed extinction for each
passband. The resulting color indices are consequently corrected from extinction.
The conversion from the Kron-Cousins to the Johnson photometric system for the optical bands
was done using the relations given by \cite{fernie1983PASP}, while the conversion between the 2MASS system
and the TCS system for the NIR bands was done using the transformation given by \cite{ramirez2005ApJ} for
dwarfs. The results are given in Table~\ref{table:ComponentTeff} and show good agreement with the
fitted effective temperatures for the two components.

\begin{table*}
\caption{$T_{\rm eff}$ for the \nsvs\ system components computed using absolute
magnitudes and color indices from the PHOEBE fitted model}             
\label{table:ComponentTeff}      
\centering          
\begin{tabular}{lcccc}     
\hline
\hline
                    &\multicolumn{2}{c}{Primary}                &\multicolumn{2}{c}{Secondary}\\
Magnitude/color &Value                  &$T_{\rm eff}$  &Value          &$T_{\rm eff}$\\
\hline
\hline
\multicolumn{5}{c}{\cite{mamajek2015}}\\
\hline
$M_V$           &7.802$\pm$0.016        &4229$\pm$9     &8.089$\pm$0.017        &4131$\pm$16\\
$M_J$           &5.388$\pm$0.025        &4252$\pm$24    &5.549$\pm$0.026        &4189$\pm$39\\
$M_K$           &4.632$\pm$0.022        &4263$\pm$26    &4.764$\pm$0.023        &4189$\pm$16\\
$V-R_C$         &0.724$\pm$0.023        &4332$\pm$60    &0.764$\pm$0.023        &4220$\pm$73\\
$V-I_C$         &1.394$\pm$0.023        &4275$\pm$40    &1.487$\pm$0.024        &4189$\pm$25\\
$V-K_S$         &3.222$\pm$0.029        &4195$\pm$9     &3.376$\pm$0.029        &3988$\pm$46\\
\hline
Average         &                       &4258$\pm$46    &                       &4151$\pm$85\\
\hline
\multicolumn{5}{c}{\cite{huang2015}}\\
\hline
$V-R_J$         &1.033$\pm$0.028        &4300$\pm$130   &1.073$\pm$0.029        &4230$\pm$120\\
$R_J-I_J$       &0.756$\pm$0.035        &4180$\pm$140   &0.809$\pm$0.035        &4080$\pm$130\\
$V-I_J$         &1.789$\pm$0.031        &4250$\pm$110   &1.882$\pm$0.032        &4160$\pm$110\\
$V-J$           &2.451$\pm$0.048        &4100$\pm$92    &2.577$\pm$0.048        &4014$\pm$89\\
$V-K$           &3.158$\pm$0.045        &4181$\pm$83    &3.313$\pm$0.045        &4085$\pm$81\\
\hline
Average         &                       &4202$\pm$76    &                       &4114$\pm$83\\
\hline
\hline
Total average   &                       &4232$\pm$65    &                       &4134$\pm$82\\
\hline
\end{tabular}
\end{table*}

We also computed the individual $T_{\rm eff}$ using the method of \cite{ribas1998}, adopting the 
V apparent magnitudes computed from the luminosity ratios ($V_1=13.461\pm0.007$, $V_2=13.749\pm0.007$), the Gaia parallax
($\pi=7.571\pm0.042$ mas), the interstellar extinction in the $V$ band ($A_V=0.063\pm0.007$), and the BC computed from
$T_{\rm eff}$ in Table~\ref{table:LCresults} ($BC_1=-0.83\pm0.10$, $BC_2=-0.96\pm0.12$).
They result in $T_{\rm eff1}=4160\pm100$ K and $T_{\rm eff2}=4070\pm110$ K, slightly lower but within $1\sigma$ agreement
with the values and uncertainties from the fit.

\begin{figure*}
\resizebox{\hsize}{!}
{\includegraphics[clip]{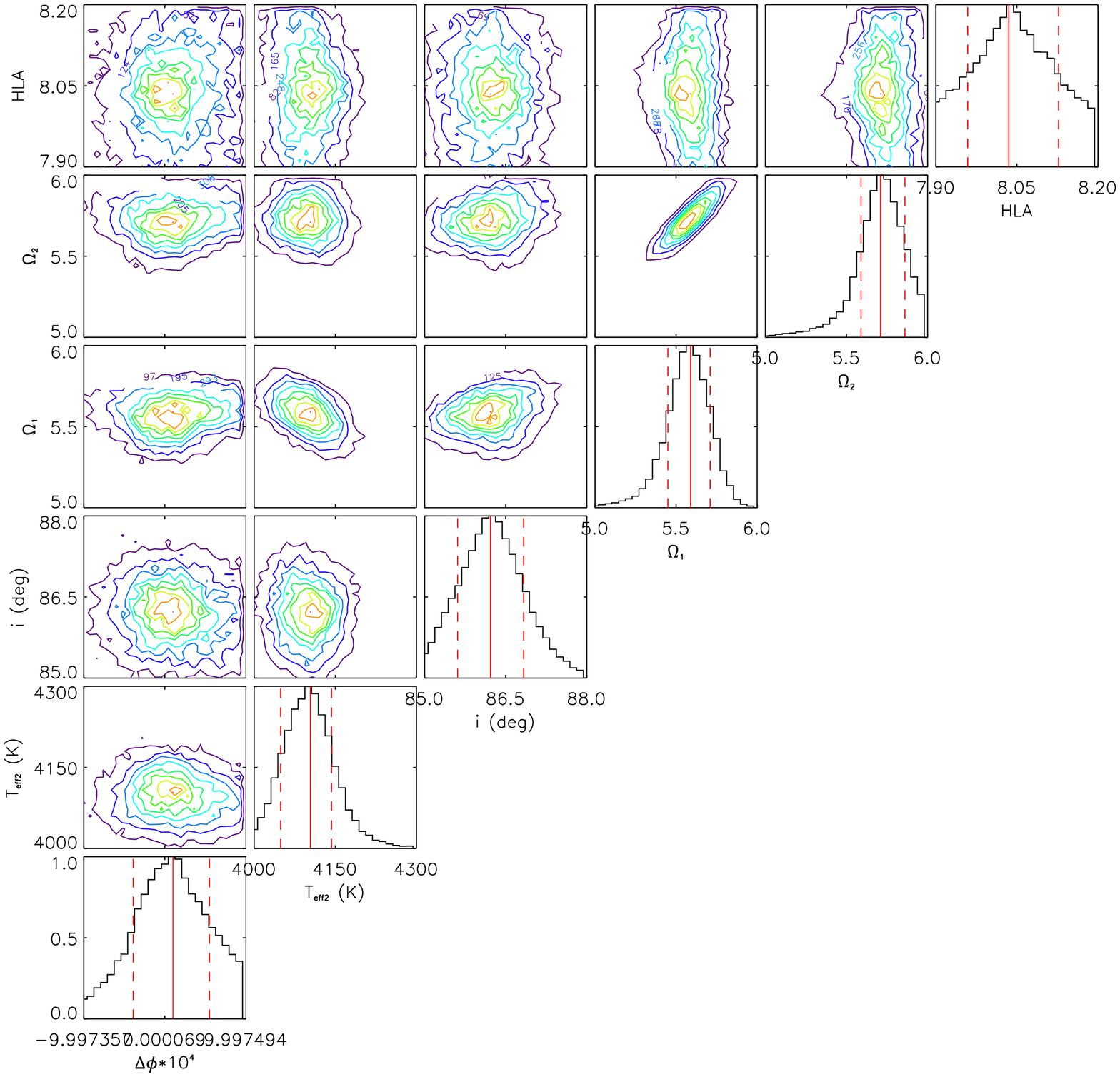}}
  \caption{Parameter correlations resulting from the MCMC fit and histograms of individual parameter distributions.
  The red vertical lines show the values adopted in this work from the maximum of the histograms.
  Dashed vertical lines indicate the 68.3\% confidence intervals, which are the adopted uncertainties in the parameters.}
\label{fig:corner}
\end{figure*}

The values of the uncertainties in the Table~\ref{table:LCresults} are the formal ones of the PHOEBE fit.
To compute robust estimations of the uncertainties for the fitted parameters we used a Markov Chain Monte-Carlo (MCMC)
wrapper for PHOEBE \citep{prsa2016}. We show the results of the computation at the bottom of Table~\ref{table:LCresults}.
Figure~\ref{fig:corner} shows the parameter correlations from MCMC simulations and histograms
of individual parameter marginalized distributions. The correlation between the
two potentials $\Omega_1$ and $\Omega_2$ is clear and related to the problem with the determination of $r_2/r_1$
mentioned before.

To ensure the robustness of the MCMC fitted light-curve model we ran
Gelman-Rubin diagnostics \citep{gelman1992} on this group of parameters.
In this test, a value near $\hat{R}\simeq$1.0 indicates a robust solution of the MCMC chain.
Table~\ref{table:gelmanrubintest} shows the $\hat{R}$ values for each fitted parameter. 

\begin{table}
\caption{Computed $\hat{R}$ values from the Gelman-Rubin test for the fitted parameters.}  
\label{table:gelmanrubintest}      
\centering                          
\begin{tabular}{l c c}        
\hline\hline                 
Fitted parameter        &$\hat{R}$\\    
\hline                                   
$\Delta\phi$            &1.00366\\
$T_{\rm eff2}$          &1.00238\\
$i$                     &1.00189\\
$\Omega_1$              &1.00187\\
$\Omega_2$              &1.00164\\
HLA\tablefootmark{a}    &1.00732\\
\hline                                   
\end{tabular}
\tablefoot{
\tablefoottext{a}{PHOEBE primary luminosity level in $V$ band.}}
\end{table}

The most striking characteristic of this system is the inverted relation among the radii and the masses for the
two components. Having very similar masses, the primary star is slightly less massive, but is larger and hotter.
Based on the resulting spot configuration and on the similar intensity of emission of the CaII H and K lines, the two
components display similar levels of activity, and so any radius inflation would be expected to be similar for the two stars. Also, radius inflation usually comes together
with temperature suppression, meaning that if a star is inflated, it is also
typically too cool. However, this is not seen in the results of the PHOEBE model for the primary.
A primary larger than the secondary is seen in all our PHOEBE fits and in the MCMC simulation, and is also confirmed
by the solutions of \cite{coughlin2007}, \cite{wolf2010}, \cite{zhang2014}, and \cite{zhang2015}, the latter
with independent data sets from ours.
As a final test, we also independently fitted with PHOEBE the $VR_C I_C$ light curves of \cite{zhang2014}
jointly with our RV results. For these light curves only a spot is needed over the primary to get a reasonable fit.
The resulting model is again compatible with no third light, and it results in potentials
$\Omega_1=5.483\pm0.014$ and $\Omega_2=5.716\pm0.016$ which translates again to $r_1>r_2$.

\section{The system of \nsvs}
\label{sec:system}

\subsection{Absolute parameters}
\label{subsec:absparams}

The absolute parameters for the \nsvs\ system are shown in Table~\ref{table:AbsDimensions}.
They were computed from the results in Tables~\ref{table:RVresults} and \ref{table:LCresults},
adopting the $T_{\rm eff2}$, potentials $\Omega_1$ and $\Omega_2$ , and orbital inclination
with their uncertainties from the MCMC computation. The volumetric radii were computed solving Eq. 1 and 2 of \cite{wilson1979}. 
The adopted uncertainty in $T_{\rm eff2}$ was taken to be the combination of the MCMC uncertainties and the
$T_{\rm eff1}$ uncertainty added in quadrature, the latter arising from the photometric colors and the empirical calibrations.
For the solar values we used the recommended International Astronomical Union (IAU) values $T_{\rm eff\sun}$=5772 K, $\log g_\sun$=4.438,
$M_{bol\sun}$=4.74 \citep{IAU_B2_2015,IAU_B3_2015}. The bolometric corrections (BC) were computed using the
BC scale by \cite{flower1996} with the corrections given in Sect. 2 of \cite{torres2010}.

{\renewcommand{\arraystretch}{1.2}
\begin{table}
\caption{\label{table:AbsDimensions}Absolute dimensions and main physical parameters of the
\nsvs\ system components.}
\centering
\begin{tabular}{lccc}
\hline\hline
Parameter                       &Primary                        &Secondary\\
\hline
Spectral type                   &K6V                            &K7V\\
$M$ ($M_\sun$)                  &0.6402$\pm$0.0052              &0.6511$\pm$0.0052\\
$R$ ($R_\sun$)          &0.687$^{+0.017}_{-0.024}$      &0.672$^{+0.018}_{-0.022}$\\
$\log g$ (cgs)                  &4.570$^{+0.022}_{-0.031}$      &4.597$^{+0.024}_{-0.029}$\\
$T_{\rm eff}$ (K)               &4240$\pm$100                   &4104$^{+107}_{-114}$\\
$(v_r \sin i)_{obs}$ (km s$^{-1}$) &67$\pm$3                    &66$\pm$3\\
$v_{sync}\sin i$ (\kms)\tablefootmark{a} &61.8$^{+1.5}_{-2.2}$  &60.5$^{+1.6}_{-2.0}$\\
$BC_V$ (mag)                    &$-0.83\pm0.10$                 &$-0.98\pm-0.14$\\
$L/L_\sun$                      &0.137$^{+0.015}_{-0.016}$      &0.115$^{+0.014}_{-0.015}$\\
$M_{bol}$ (mag)                 &6.89$^{+0.12}_{-0.13}$         &7.08$^{+0.13}_{-0.14}$\\
$M_V$ (mag)                     &7.72$\pm$0.16          &8.06$^{+0.19}_{-0.20}$\\
$i$ (deg)                       &\multicolumn{2}{c}{86.22$\pm$0.61}\\
$a$ ($R_\sun$)                  &\multicolumn{2}{c}{3.1149$\pm$0.0079}\\
$M_{Vtot}$ (mag)                &\multicolumn{2}{c}{7.12$^{+0.12}_{-0.13}$}\\
$d$ (pc)                        &\multicolumn{2}{c}{135.2$^{+7.6}_{-7.9}$}\\
\hline
\end{tabular}
\tablefoot{
\tablefoottext{a}{Projected rotational velocity expected for synchronous rotation and
a circular orbit.}
}
\end{table}
}

Based on their effective temperatures, the two stars have spectral types K6V and K7V \citep{mamajek2015}.
Given the uncertainties in the $T_{\rm eff}$ we adopted uncertainties of $\pm$1 for the spectral types.
The masses both have  relative uncertainties of $\sigma(M)/M\simeq\pm$0.8\%, while the radii have (+2.5\%, $-$3.5\%) for $R_1$,
and (+2.7\%, $-$3.3\%) for $R_2$, which makes this system suitable for testing the stellar models.
The departure of the two stars from a spherical model is very small with the two stars well inside
their Roche lobes.

The physical parameters of this system suggest that the times
for the tidal synchronization and orbital circularization to occur are very short \citep{hut1981}.
Following \cite{hilditch2001}, these times, in units of years, can be computed as

\begin{equation}\label{eq:tsync}
t_{sync}\simeq10^4\left(\frac{1+q}{2q}\right)^2P^4
,\end{equation}

\begin{equation}\label{eq:tcirc}
t_{circ}\simeq10^6q^{-1}\left(\frac{1+q}{2}\right)^{5/3}P^{16/3},
\end{equation}

where $q=M_2/M_1$ is the mass ratio (from Table~\ref{table:RVresults}) and $P$ is the
orbital period in days. For \nsvs\ these times are $t_{sync}\simeq10^3$ yr and $t_{circ}\simeq4.5\times10^4$ yr.
These values are small compared to typical ages of low-mass stars, and justify our assumptions about
the synchronicity parameters in Sect.~\ref{subsec:rotation}.

\subsection{Age, distance, and space velocities}
\label{subsec:age}

Using the bolometric luminosities of the two components in Table~\ref{table:AbsDimensions}, and bolometric
corrections ($BC_V$), we obtain a combined absolute V magnitude for the system of $M_{Vtot}=7.12^{+0.12}_{-0.13}$.
Using the visual apparent magnitude of the system ($V=12.843\pm0.005$, see Table~\ref{table:calibrated_phot})
and the interstellar reddening in the $V$ band, this results in a distance modulus of $m-M=5.72^{+0.12}_{-0.13}$,
which translates into a distance of $d=135.2^{+7.6}_{-7.9}$ pc.
The computed distance to the system is in excellent agreement with the Gaia value ($d_{Gaia}=131.6\pm0.8$ pc)
with the bulk of the uncertainty coming from the $T_{\rm eff}$ and the bolometric corrections that arise from them.

We also computed the Galactic $(U,V,W)$\footnote{Positive values of $U$, $V$, and $W$ indicate velocities
toward the Galactic center, Galactic rotation, and north Galactic pole, respectively.}
space velocities applying the prescription of \cite{johnson1987},
the  systemic RV of the system (see Sect.~\ref{subsec:RV_fit}), and the Gaia distance and
proper motion measurements: $\mu_\alpha=-37.938\pm0.052$ mas yr$^{-1}$, $\mu_\delta=-23.381\pm0.051$ mas yr$^{-1}$.
The resultant velocities are $U=-10.07\pm0.21$ \kms, $V=-30.21\pm0.18$ \kms, and $W=-0.04\pm0.22$ \kms,
and a total space velocity of $S=31.84\pm0.35$ \kms.
These Galactic velocities indicate thin disk kinematics for \nsvs\, as the $W$ velocity is nearly zero.

In an attempt to constrain the age of the system we checked a number of kinematic criteria with little success.
This binary is outside the area defined by \cite{eggen1989} as belonging to the young Galactic disk, and
the velocities lie just over the $V$ boundary of the criteria of \cite{leggett1992}
($-50<U<20$, $-30<V<0$, $-25<W<10$, all in \kms); also the binary is not within any known early-type
or late-type population tracer \citep{skuljan1999}, and cannot be related to any known
moving group \citep{montes2001,maldonado2010}. Therefore, we cannot
impose constraints on the age of the system based on kinematic criteria.
In addition, the short period of this system cannot impose constraints based on the synchronization
or the circularization times. At best, all we can say from this system, based on a solar or sub-solar metallicity
and the typical age of K stars, is that the system would be a main sequence star with an undefined age of one or more gigayears.

\begin{figure} 
\centering
\includegraphics[width=\hsize]{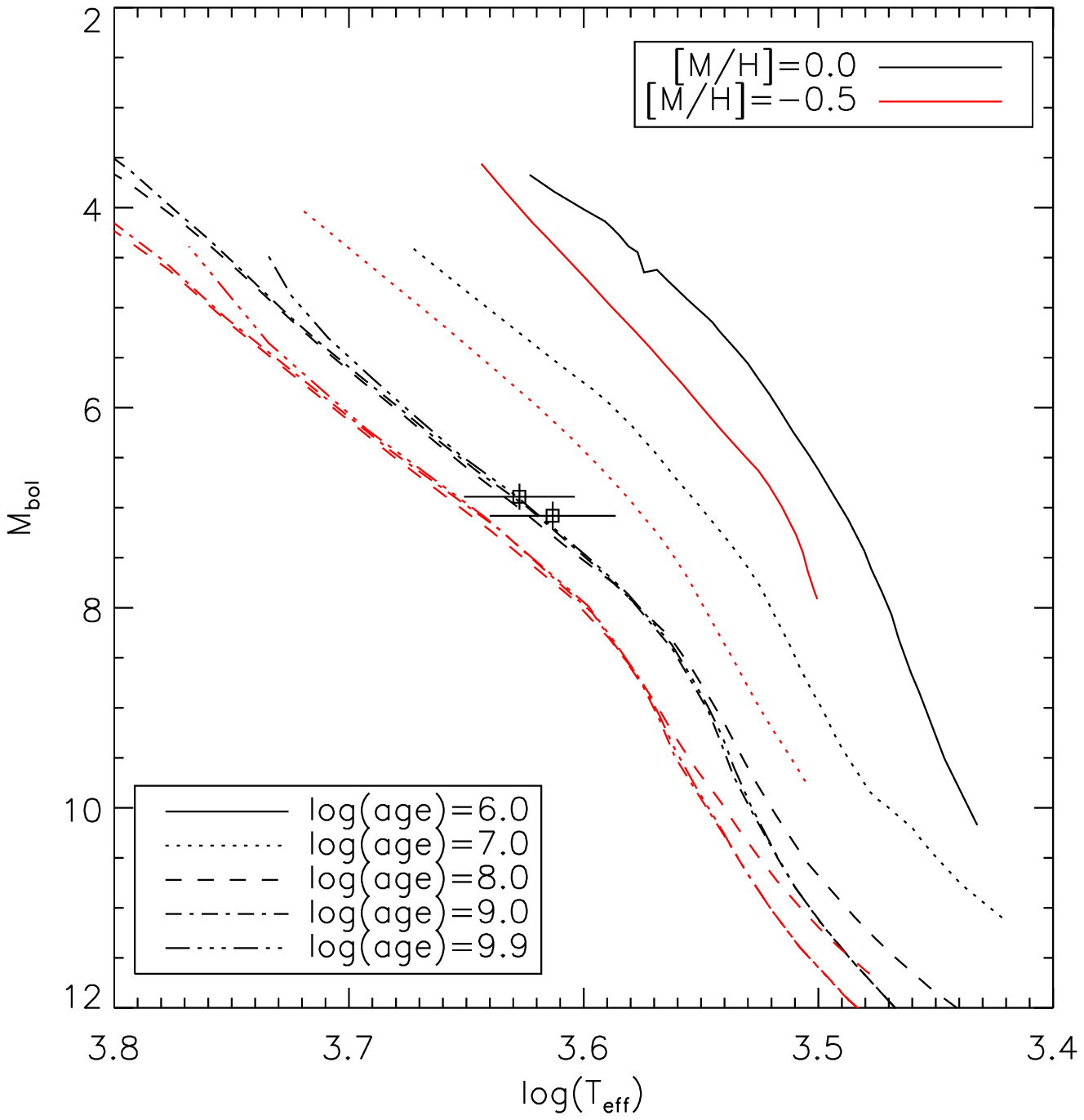}
  \caption{Position in the $M_{bol}$-$\log T_{\rm eff}$ diagram of the \nsvs\ system components.
  BCAH98 isochrones are for $[M/H]=0.0$ (black) and $[M/H]=-0.5$ (red) for $\log(age)$=6.0, 7.0, 8.0, 9.0 and 9.9.
  The ages of the components are compatible with $\log(age)$=8.0 and older or even with younger ages
  if the metallicity is lower than solar.}
\label{fig:Mbol_logTeff}
\end{figure}

At last, Fig.~\ref{fig:Mbol_logTeff} shows a $M_{bol}$-$\log T_{\rm eff}$ diagram of the two components
of this system with the \cite{baraffe1998} (hereafter BCAH98) isochrones overplotted.
Since the metallicity of \nsvs\ is unknown we included isochrones for solar metallicity $[M/H]=0.0$ dex
represented as black lines, and for low metallicity with $[M/H]=-0.5$ dex, represented by red lines.
Assuming solar metallicity, all that can we say from that figure is that this system is older than
$\log(age(yr))\sim8$. Any age of $\log(age(yr))\sim8$ or older fits the two components equally
well.
In this scenario, \nsvs\ would be a main sequence system.
If low metallicity is assumed, for example $[M/H]=-0.5$ dex, the age of the system could be fitted by a younger
isochrone, with an age halfway $\log(age(yr))=7-8$. Metallicities slightly over $[M/H]=-0.5$
could also fit an old system given the uncertainties in the position of the components
in the $M_{bol}$--$\log T_{\rm eff}$ plane. Specific metallicity measurements for this system would
help to resolve this ambiguity; though as mentioned before they would be difficult to measure
because of their high rotation speeds.

\subsection{Comparison with models}
\label{subsec:comparison_with_models}

\begin{figure}
\centering
\includegraphics[width=\hsize]{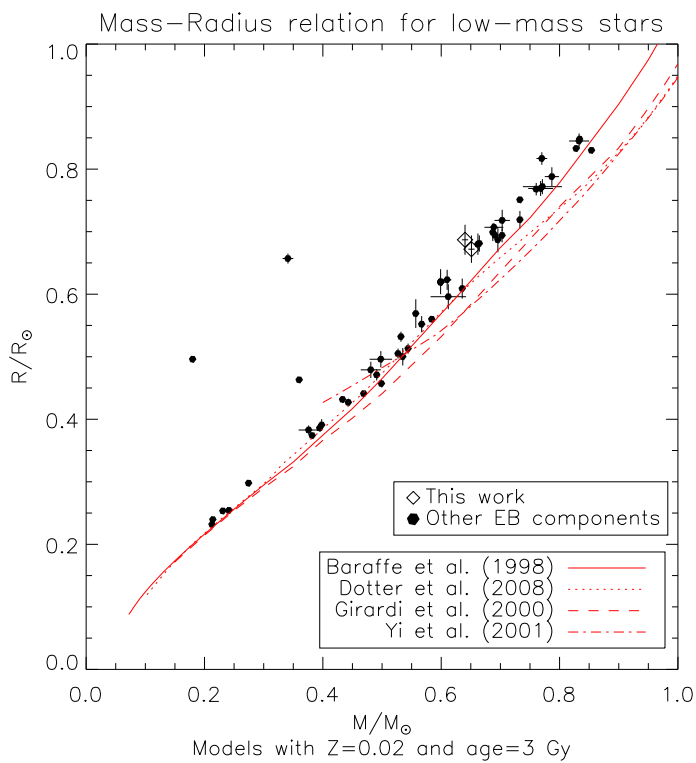}
  \caption{Mass--radius relations of stars between 0.2 and 1.0 $M_\sun$ predicted by four
  stellar models (see text). The filled circles represent well-known LMDEBs used as benchmarks
  for the models. The diamonds represent the location of \nsvs\ system components. The uncertainties in the
  parameters are represented by bars, though in many of the systems these are smaller than the symbols.
  Metallicity and age are set for reference of the whole set of stars and are not a fit to
  the \nsvs\ parameters.} \label{fig:M-R}
\end{figure}

\begin{figure}
\centering
\includegraphics[width=\hsize]{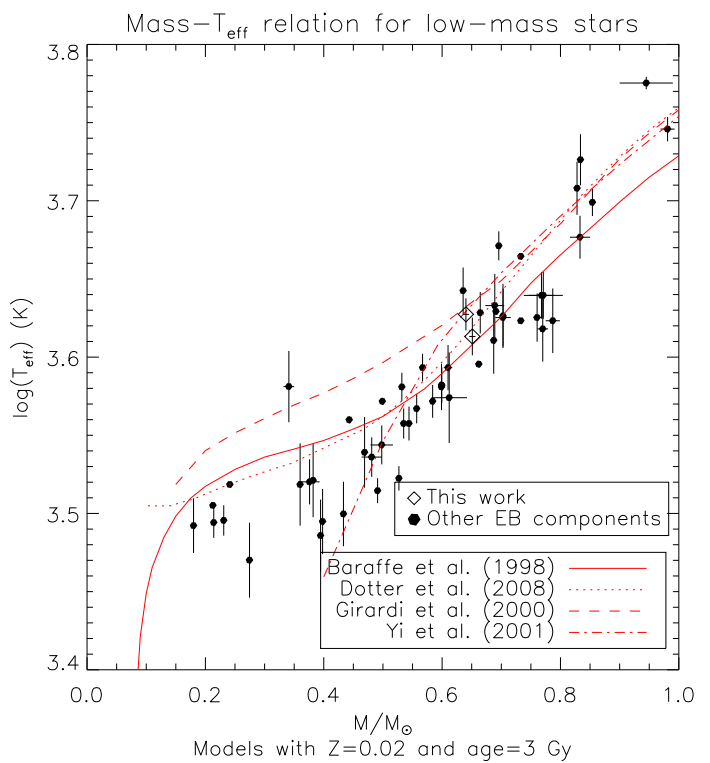}
  \caption{Mass--$\log T_{\rm eff}$ relations for the same LMDEB systems as in Fig.~\ref{fig:M-R}.
  Metallicity and age are set for reference
  of the whole set of stars and  are not a fit to the \nsvs\ parameters.} \label{fig:M-Teff}
\end{figure}

In Fig.~\ref{fig:M-R} we plot the masses and radii for the \nsvs\ components and other benchmark LMDEBs
collected from the literature. The individual components of benchmark LMDEBs are plotted as filled
circles and the components of \nsvs\ are plotted as open diamonds. All of them are plotted with their
uncertainties, but in many cases those are smaller than the plot symbols. It must be
stressed that some of the systems taken from the literature quote only formal uncertainties and that actual
error bars should be larger. The complete list of systems plotted can be found in \cite{iglesias-marzoa2017}
(see their Table~15) from which we selected those with relative uncertainties of less than 5\% in mass and
radius.

We also plotted four theoretical models which predict mass--radii relations for the
stellar low-mass regime, namely the models of \cite{baraffe1998}, \cite{dotter2008},
\cite{girardi2000}, and \cite{yi2001}. The selected models have solar metallicity (Z=0.02) and
an age of 3 Gyr, and they must be taken only as reference and for comparison with the sample of objects.
They are not fits to the components of \nsvs.
For the \cite{baraffe1998} model we selected a mixing length parameter of $\alpha=l/H_p=1$.

The components of \nsvs\ follow the same trend seen in other LMDEBs in their range of masses, that is,
the components are oversized with respect to the radius predicted by the models: the radius of
the primary is about 15\% larger than the model predicts, and the radius of the secondary is
about 12\% larger. Those differences are clearly larger than the computed radii uncertainties.

Figure~\ref{fig:M-Teff} shows the mass--$T_{\rm eff}$ relation for the same LMDEB systems plotted
in Fig.~\ref{fig:M-R} and for \nsvs. In this case, the components of \nsvs\ are well represented
by the stellar models; in particular, the Dartmouth model \citep{dotter2008} passes between
the two components.

\section{Conclusions}
\label{sec:conclusions}

We present a set of reliable physical parameters of the \nsvs\ system components based on optical and IR
differential photometry and on spectroscopic RV observations.
This LMDEB system is composed of two oversized main sequence K stars. The resulting physical parameters
for the primary star are
$M_1=0.6402\pm0.0052$ $M_\sun$, $R_1=0.687^{+0.017}_{-0.024}$ $R_\sun$, and $T_{\rm eff1}=4240\pm100$ K.
For the secondary, the parameters are
$M_2=0.6511\pm0.0052$ $M_\sun$, $R_2=0.672^{+0.018}_{-0.022}$ $R_\sun$, and $T_{\rm eff2}=4104^{+107}_{-114}$ K.
The uncertainties in mass and radius were derived in a robust way using MCMC and are at the level
of $\sim0.8\%$ for mass and $\sim3\%$ for radius, allowing for comparison with current  models of low-mass stars.
As seen for other LMDEBs, the components show inflated radii, and $T_{\rm eff}$ depression is found at a similar level to that of other LMDEBs.

The orbit is circular with $i=86.22\pm0.61$ degrees and our derived distance is in excellent agreement
with the value derived from Gaia parallax. This imposes a strict constraint on the luminosity of the
system, and, as a result, on the effective temperature.
We do not detect any hint of third light in our numerous tests on the LCs of this system.
As a consequence, the period change reported by \cite{zhang2015} is unlikely to be explained by the presence
of a third main sequence body in the system, though a white dwarf or a substellar object is still possible.
Both components show signs of activity in the form of spots and CaII H and K emission lines
which complicates the analysis of the LCs.
The age of the system cannot be established using isochrones or kinematical properties.
Therefore, although unlikely
given the typically old age
of the stars in this mass range, a young system cannot be completely discarded.

\begin{acknowledgements}

We thank the anonymous referee for their careful reading of the manuscript and helpful comments.
This article is based on observations made with the Carlos S\'anchez IR telescope (TCS)
and the IAC80 optical telescope operated on the island of Tenerife by the Instituto de
Astrofisica de Canarias in the Spanish Observatorio del Teide. Also is based on observations
made with the Tillinghast Reflector Echelle Spectrograph (TRES) on the 1.5-meter
Tillinghast telescope at the Smithsonian Astrophysical Observatory’s Fred L. Whipple Observatory.
GT acknowledges partial support from the NSF through grant AST-1509375.
RIM acknowledges support through the Programa de Acceso a Instalaciones Cientificas Singulares (E/309290).
IRAF is distributed by the National Optical Astronomy Observatory, which is operated
by the Association of Universities for Research in Astronomy (AURA) under a cooperative
agreement with the National Science Foundation. 
This research has made use of the SIMBAD database, operated at CDS, Strasbourg, France,
and of NASA's Astrophysics Data System Bibliographic Services.
Also, it used data products from the Two Micron All Sky Survey, which is a
joint project of the University of Massachusetts and the Infrared Processing and
Analysis Center, California Institute of Technology, and is funded by NASA and the
National Science Foundation. We acknowledge the BRNO database for publishing their observations,
and in particular to Katerina Honkova for her kind reply to our questions.

\end{acknowledgements}

%
%

\bibliographystyle{aa} 

\bibliography{master} 

\end{document}